\documentclass[twocolumn,showpacs,prc,superscriptaddress]{revtex4-1}
\usepackage{graphicx}
\usepackage{dcolumn}
\usepackage{bm}

\begin{document}

\title{Fission fragment mass distribution in $^{210}$Po and $^{213}$At }

\author{A. Sen} 
\affiliation{Variable Energy Cyclotron Centre,  1/AF, Bidhan  Nagar,
  Kolkata  700064, India}
	
	\affiliation{Homi Bhabha National Institute, Training School Complex, Anushakti Nagar, Mumbai - 400094, India}
\author {T. K. Ghosh} \email{E-mail:tilak@vecc.gov.in}
\author{S. Bhattacharya} \thanks{Raja Ramanna Fellow}
\author{K. Banerjee} \thanks{Present address:Department of Nuclear Physics, Research School
of Physics and Engineering, Australian National University, Canberra,
ACT 2601, Australia}
\author{C. Bhattacharya}
\author {S. Kundu}
\author{G. Mukherjee}
\author {A. Asgar}
\affiliation{Variable Energy Cyclotron Centre,  1/AF, Bidhan  Nagar,
  Kolkata  700064, India}
	
	\affiliation{Homi Bhabha National Institute, Training School Complex, Anushakti Nagar, Mumbai - 400094, India}
\author {A. Dey}
\author{A. Dhal}
\affiliation{Variable Energy Cyclotron Centre,  1/AF, Bidhan  Nagar,
  Kolkata  700064, India}
	
\author{Md. Moin Shaikh}
\affiliation{Saha Institute of Nuclear Physics,  1/AF,  Bidhan  Nagar,
  Kolkata  700064, India.}
\author {J.K. Meena}
\affiliation{Variable Energy Cyclotron Centre,  1/AF, Bidhan  Nagar,
  Kolkata  700064, India}
\author {S. Manna}
\author {R. Pandey}
\affiliation{Variable Energy Cyclotron Centre,  1/AF, Bidhan  Nagar,
  Kolkata  700064, India}
	
	\affiliation{Homi Bhabha National Institute, Training School Complex, Anushakti Nagar, Mumbai - 400094, India}
	
\author {T.K. Rana}
\affiliation{Variable Energy Cyclotron Centre,  1/AF, Bidhan  Nagar,
  Kolkata  700064, India}
\author {Pratap Roy}
\author {T. Roy} 
\author {V. Srivastava} \thanks{Present address: Inter University Accelerator Centre,
Aruna Asaf Ali Marg, New Delhi-110067, INDIA.}
\affiliation{Variable Energy Cyclotron Centre,  1/AF, Bidhan  Nagar,
  Kolkata  700064, India}
	
	\affiliation{Homi Bhabha National Institute, Training School Complex, Anushakti Nagar, Mumbai - 400094, India}

	\author{P. Bhattacharya}  
	\affiliation{Saha Institute of Nuclear Physics,  1/AF,  Bidhan  Nagar,
  Kolkata  700064, India.}

\date{\today}


\begin{abstract}

\textbf{Background}: The influence of shell effect on the dynamics of the fusion fission process and it's evolution with excitation energy in the pre-actinide Hg-Pb region in general is a matter of intense research in recent years. In particular, a strong ambiguity remains for the neutron shell closed  $^{210}$Po nucleus  regarding the role of shell effect in fission around $\approx$ 30 - 40 MeV of excitation energy. 

\textbf{Purpose}:  We have measured the fission fragment mass distribution of  $^{210}$Po populated using fusion of  $^{4}$He + $^{206}$Pb  at different excitation energies and compare the result with recent theoretical predictions as well as with our previous measurement for the same nucleus populated through a different entrance channel. Mass distribution in the fission of the neighbouring nuclei $^{213}$At is also studied for comparison. 

\textbf{Methods}: Two large area Multi-wire Proportional Counters (MWPC) were used for complete kinematical measurement of the coincident fission fragments. The time of flight differences of the coincident fission fragments were used to directly extract the fission fragment mass distributions. 

\textbf{Results}: The measured fragment mass distribution for the reactions $^{4}$He + $^{206}$Pb  and $^{4}$He + $^{209}$Bi were symmetric and the width of the mass distributions were found to increase monotonically with excitation energy above 36.7 MeV and 32.9 MeV, respectively, indicating the absence of shell effects at the saddle. However, in the fission of $^{210}$Po, we find minor deviation from symmetric mass distributions at the lowest excitation energy (30.8 MeV). 

\textbf{Conclusion}: Persistence of shell effect in fission fragment mass distribution of $^{210}$Po was observed at the excitation energy $\approx$ 31 MeV as predicted by the theory; at higher excitation energy, however, the present study reaffirms the absence of any shell correction in the fission of $^{210}$Po.

\end{abstract}

\maketitle{
\begin{center}
\textbf{Introduction}
\end{center}}

In recent years, our understanding of shell effect in nuclei and it's manifestation in the fission process has come under serious scrutiny since the discovery of asymmetric fission of $^{180}$Hg at low excitation energies ($\leq$ 10 MeV) \cite{HgPRL}. Intense experimental and theoretical investigations followed, in and around the Hg region in particular, to explain the observed behaviour. Detailed theoretical studies using both macroscopic microscopic model (MMM) \cite{Moller2015} and self-consistent finite temperature density functional theory (FT-DFT) \cite{McDonnell} have indicated that, apart from the shell effects of the nascent fragments, the fission trajectory through the multi-dimensional potential energy surface plays the decisive role in deciding the exit channel asymmetry of the fission fragments. It was further emphasised that a thorough study of the Hg-Pb region, both theoretical and experimental, is crucial for proper understanding of the evolution of the interplay of shell effect and dynamics of fission with excitation energy in this region. Though some progress in this direction has recently been made so far as the detailed microscopic predictions are concerned \cite{HgPRL,McDonnell} (and references therein), experimental data are really scarce for proper validation of the theory. An overview of mass and energy distribution studies for nuclei lighter than thorium was presented in ref \cite{Gonnenwein}.

One of the few nuclei in this region which has been studied quite extensively to understand the persistence of the shell effect  near the fission saddle is $^{210}$Po \cite{Itkis,Itkis1,MulginNPA98}, a N=126 neutron closed shell system. A few recent experimental and theoretical studies indicated strong ambiguity for this system regarding the presence of shell correction at the saddle point. An anomalous increase in fission fragment angular anisotropy was reported in the fission of $^{210}$Po at excitation energy $\approx$ 40 - 60 MeV, which was conjectured as an indirect evidence of shell effect at saddle due to neutron shell closure \cite{AradhanaPRL}. However,  recent dynamical calculation using stochastic Langevin equation \cite{Schmitt} claimed that the observed angular anisotropy could be well explained with only macroscopic potential energy landscape without considering any shell effect at saddle point. Statistical model calculation \cite{Kripa15} for the nucleus $^{210}$Po, populated in light and heavy ion induced reactions, however could describe the excitation functions without the requirement of shell correction at the saddle, but required a huge fission delay to fit the pre-scission neutron multiplicity data in heavy ion induced reaction. Reanalysis \cite{Kripa02} of the  $^{210}$Po data \cite{AradhanaPRL} with inclusion of multi-chance nature of fission was found to reduce the anisotropy anomaly. 

Recent fission fragment mass distribution measurement by us  \cite{AbhirupRapid} for the fissioning nuclei $^{206,210}$Po to look for the signatures of shell correction at the saddle also produced null result. No significant deviation of mass distribution was found between $^{206}$Po and $^{210}$Po and both the distributions could be explained using realistic macroscopic potential only, indicating that there was no influence of  N=126 shell closure on the mass distribution of  $^{210}$Po above 40 MeV of excitation energy, contrary to the reported angular anisotropy result.  This however does not preclude the conjecture of the existence of shell effect at still lower energies; a more complete picture of shell effect in fission will emerge only when the investigations are extended to lowermost energies.

In recent years, detailed microscopic mapping of the multidimensional potential energy surface (PES) for $^{210}$Po at various excitation energies (E*), starting from the lowest ground state, has been performed using both MMM \cite{Moller2015} and FT-DFT \cite{McDonnell} approaches to study its effect on fission process. The FT-DFT calculation has indicated that, whereas at E* = 0 MeV, the dominant fission pathway favours a slight mass asymmetry; at  E* $\geq$ 40 MeV, the fission fragment mass distribution becomes purely symmetric in nature. In absence of any calculation at any intermediate energies, the location of the transition to symmetry could not be identified in \cite{McDonnell}. On the other hand, in the MMM approach \cite{Moller2015}, using the Brownian shape motion along the PES, systematic calculation of fragment mass distributions of $^{210}$Po, $^{213}$At and the neighbouring pre-actinide nuclei has been performed  over a wide range of excitation energies. Interestingly, this calculation indicated that there may be a small presence of the asymmetric fission in $^{210}$Po (at $\leq$ ≤ 1 $\%$ yield level) at E* $\approx$ 31 - 32 MeV. Experimental validation of this prediction is crucial for proper understanding of the shell effect in fission of $^{210}$Po.

Here, we present the results of our new complete kinematical measurement of the  fission fragment mass distributions of $^{210}$Po and  $^{213}$At down to the excitation energy $\approx$ 30 MeV to look for the predicted asymmetric fission pathway in $^{210}$Po. The compound nuclei $^{210}$Po* and  $^{213}$At* were produced through light ion induced fusion reaction $^{4}$He + $^{206}$Pb,$^{209}$Bi as it was convenient for producing compound systems at  lower excitation energy and angular momentum in general; in addition, there is reduced complication due to entrance channel dynamics (by mixture of transfer induced or non compound nuclear fission) in this case.

\maketitle{
\begin{center}
\textbf{Experimental Details}
\end{center}}

Alpha ($^{4}$He) beam of energies 37 to 55 MeV from the K-130 cyclotron at VECC, Kolkata was bombarded on enriched $^{206}$Pb  and $^{209}$Bi targets, to produce the compound nuclei of $^{210}$Po$^{*}$ and $^{213}$At$^{*}$, respectively. The target of $^{206}$Pb of thickness 250 $\mu$g/cm$^{2}$, was prepared by evaporating enriched $^{206}$Pb on a carbon backing of 20 $\mu$g/cm$^{2}$. The self supporting target of $^{209}$Bi was also prepared by evaporation technique and was of thickness 440 $\mu$g/cm$^{2}$.  The targets were set at an angle of 45$^{o}$ to the beam axis. The fission fragments were detected using two large area position sensitive multi-wire proportional chambers (MWPC) of active area 20 cm $\times$ 6 cm. The forward detector and the backward detector were centered at 60$^{o}$ and 114$^{o}$ with respect to the beam axis, respectively. The forward detector had an angular coverage of 60$^{o}$  while the backward detector covered 72$^{o}$. The angles were selected on the basis of Viola's systematics \cite{Viola} corresponding to symmetric fission fragments for complete momentum transfer of the projectile. The detectors were operated at a low pressure of 3 Torr of isobutane gas so that  elastic and quasi-elastic particles were transparent in the detectors. The time of flights of the fission fragments, the positions of impact of the fission fragments with the detector and the energy loss of the fission fragments in the gas volume of the detectors were recorded, event by event basis on a VME based data acquisition system. Counts from the Faraday cup were used for beam flux monitoring and normalization of the data.

\maketitle{
\begin{center}
\textbf{Analysis of data}
\end{center}}

The folding angle distribution for all the fission fragments were constructed and the peak was seen to conform with that associated with complete transfer of momentum of the projectile. Since the projectile was $^{4}$He beam, transfer induced fusion-fission was not expected in this reaction. This was evident by the symmetric shape of the folding angle distribution. The correlated fission fragments were clearly separated in the time correlation (TOF-TOF) and the energy loss spectra. The masses of the fission fragments were calculated using the time of flight difference between the correlated fragments, the azimuthal and polar angles, the recoil velocities and the momenta of the fission fragments. This prescription for calculation for masses is described in \cite{TilakdaNIM}. In order to correct for the energy loss of the fragment in the backing (for the $^{206}$Pb target) and half the target thickness, the gold deconvolution \cite{deconv1} process was implemented to find the correct width of the mass distributions.

\maketitle{
\begin{center}
\textbf{Results and discussion}
\end{center}}

\begin{figure}
\includegraphics*[scale=0.38, angle=0]{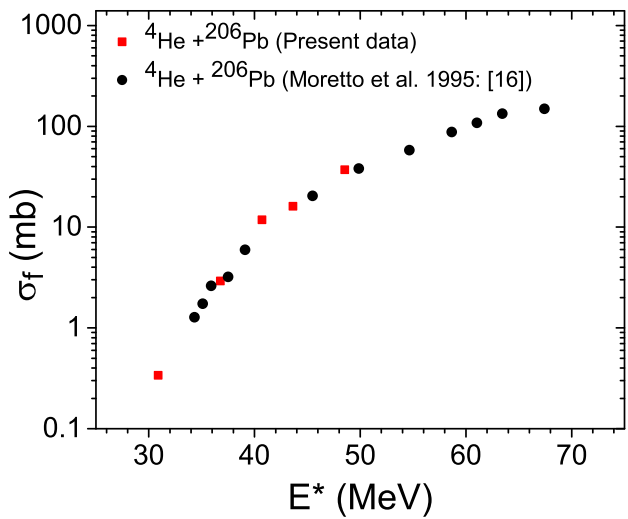}
\caption{\label{fig:fig1}~(Color online) Measured fission cross section for the reaction $^{4}$He + $^{206}$Pb. The red points correspond to the presently measured values, and the black points correspond to previously measured values of Moretto el al. \cite{MorettoPRL95}. The uncertainties in the cross sections are smaller than the points.}
\end{figure}

\begin{figure}
\includegraphics*[scale=0.3, angle=0]{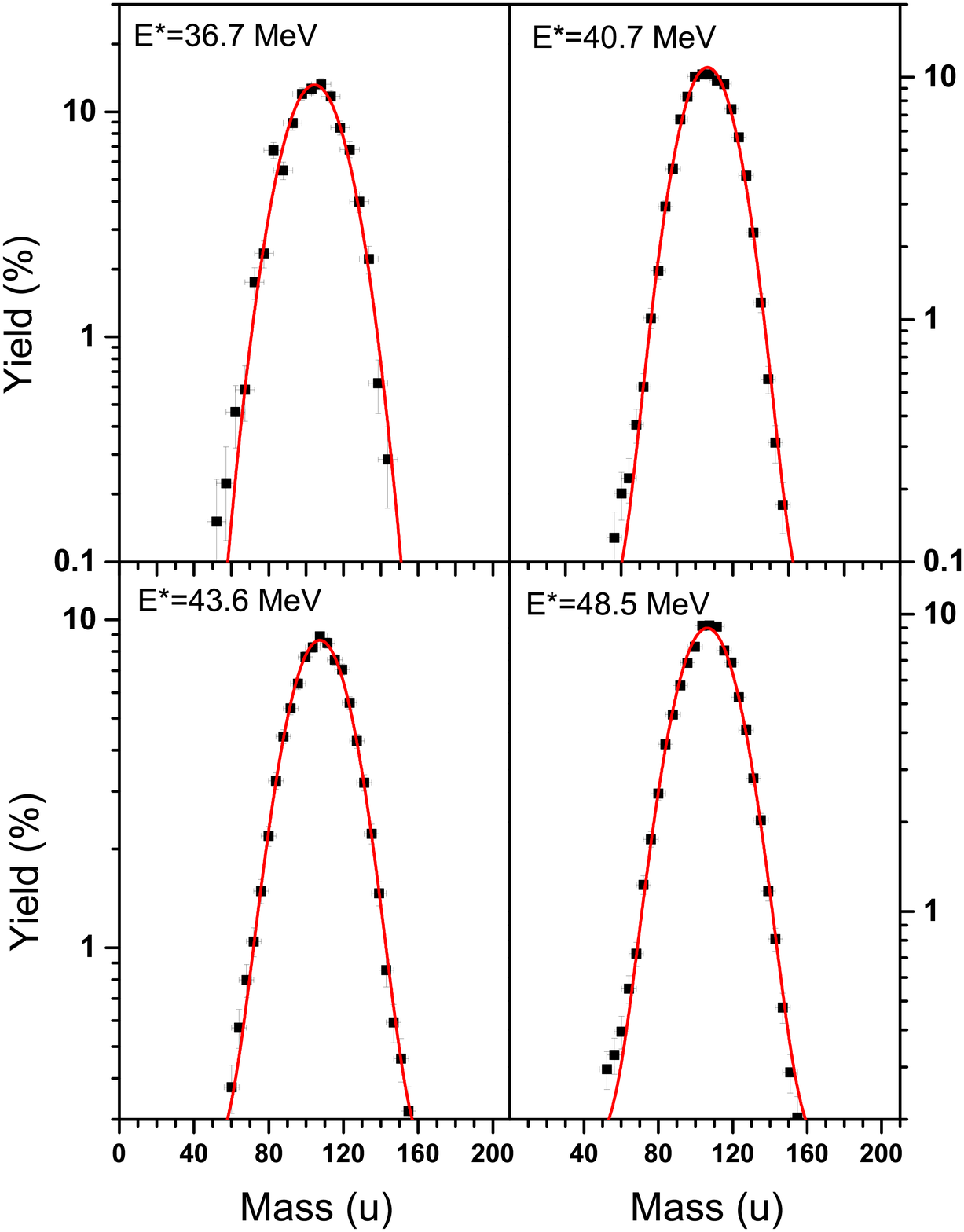}
\caption{\label{fig:fig2}~(Color online) Fission fragment mass distributions for $^{210}$Po. The red line indicates the single Gaussian fit.}
\end{figure}
 
The measured fission cross section for the reaction $^{4}$He + $^{206}$Pb is shown in Figure 1. This was calculated using the total number of counts recorded and normalized with the angular coverage of the detectors, time of collection of the events and the Faraday cup readings. The cross sections thus obtained were compared with the previously reported data of L.G. Moretto et al \cite{MorettoPRL95}, and found to follow the same trend. The error in the fission cross section due to limited angular coverage of our detectors was taken into account in addition to the statistical error while calculating uncertainty in the measured cross section. The overall uncertainty was $\approx$ 7$\%$. 

\begin{figure}
\includegraphics*[scale=0.4, angle=0]{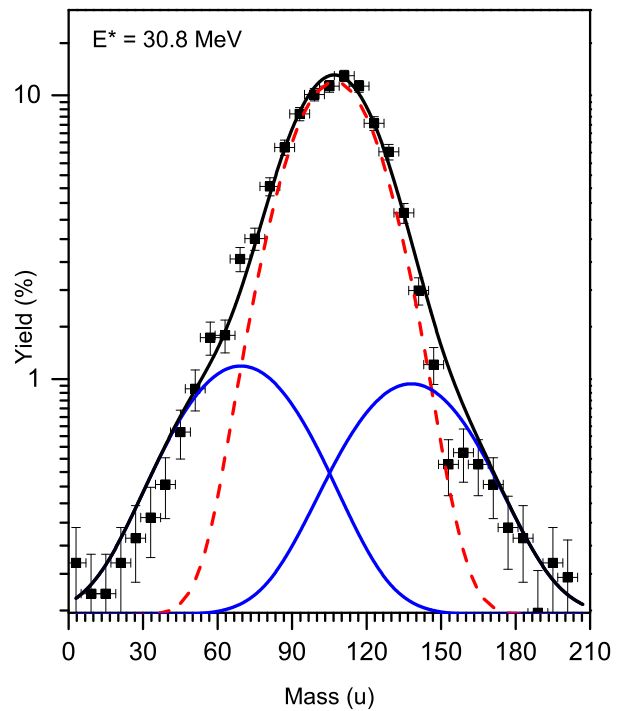}
\caption{\label{fig:fig3}~(Color online) Fission fragment mass distribution of $^{210}$Po at an excitation energy of 30.8 MeV. The red dashed line indicates the Gaussian fit corresponding to the symmetric fragments, while the blue solid lines show the asymmetric components. The overall fitting is shown by the solid black line. }
\end{figure}

Figure 2 shows the fission fragment mass distributions for the reaction $^{4}$He + $^{206}$Pb populating the nucleus $^{210}$Po at various excitation energies. The shape of the fission fragment mass distribution is symmetric and could be well fitted with a single Gaussian. No significant deviation at the tails of the Gaussian  indicating presence of asymmetric fission pathway could be identified in any of the distributions. This may be indicative of the fact that the shell effects may not play a role in determining the dynamics of fusion fission of $^{210}$Po at the excitation energies of 36.7 MeV and above. Theoretical predictions for this system at these energies \cite{Moller2015} are also consistent with the present observations.

However, the observed mass distribution for fission of $^{210}$Po at the lowest excitation energy of 30.8 MeV, shown in  figure 3, is clearly different from those observed at higher energies. It is seen that, in this case too, symmetric component dominates the mass distribution, which is represented by a single Gaussian. However, deviation from the Gaussian in the tail regions is clearly evident from the graph. The mass distribution could be best (chi square) fitted by three Gaussians; a dominant Gaussian peaked at the symmetric mass and two other Gaussians, one on the low mass side (peaking at $\approx$ 70) and another on the high mass side (peaking at $\approx$ 132). This is qualitatively consistent with the theoretical predictions of Moller et al. \cite{Moller2015} at the excitation energy of 31.43 MeV, where a weak asymmetric component ($\approx$ 1 $\%$ yield level) was shown to coexist along with the strong symmetric mass distribution. The heavier mass side of the asymmetric distribution was predicted to peak around $\approx$ 132 (doubly magic $^{132}$Sn), a clear indication of the persistence of the shell effect of the nascent fragments. Interestingly, the heavier mass peak of the asymmetric component calculated from the present data, within the limits of experimental resolution, closely matches  with the theoretical value. Therefore, this observed deviation from the single Gaussian and the presence of the side wings in the data may be attributed to the influence of shell effects in the dynamics governing fusion fission and it can be said that some shell effect survives  for $^{210}$Po at least up to an excitation energy of $\approx$ 31 MeV.

\begin{figure}
\includegraphics*[scale=0.3, angle=0]{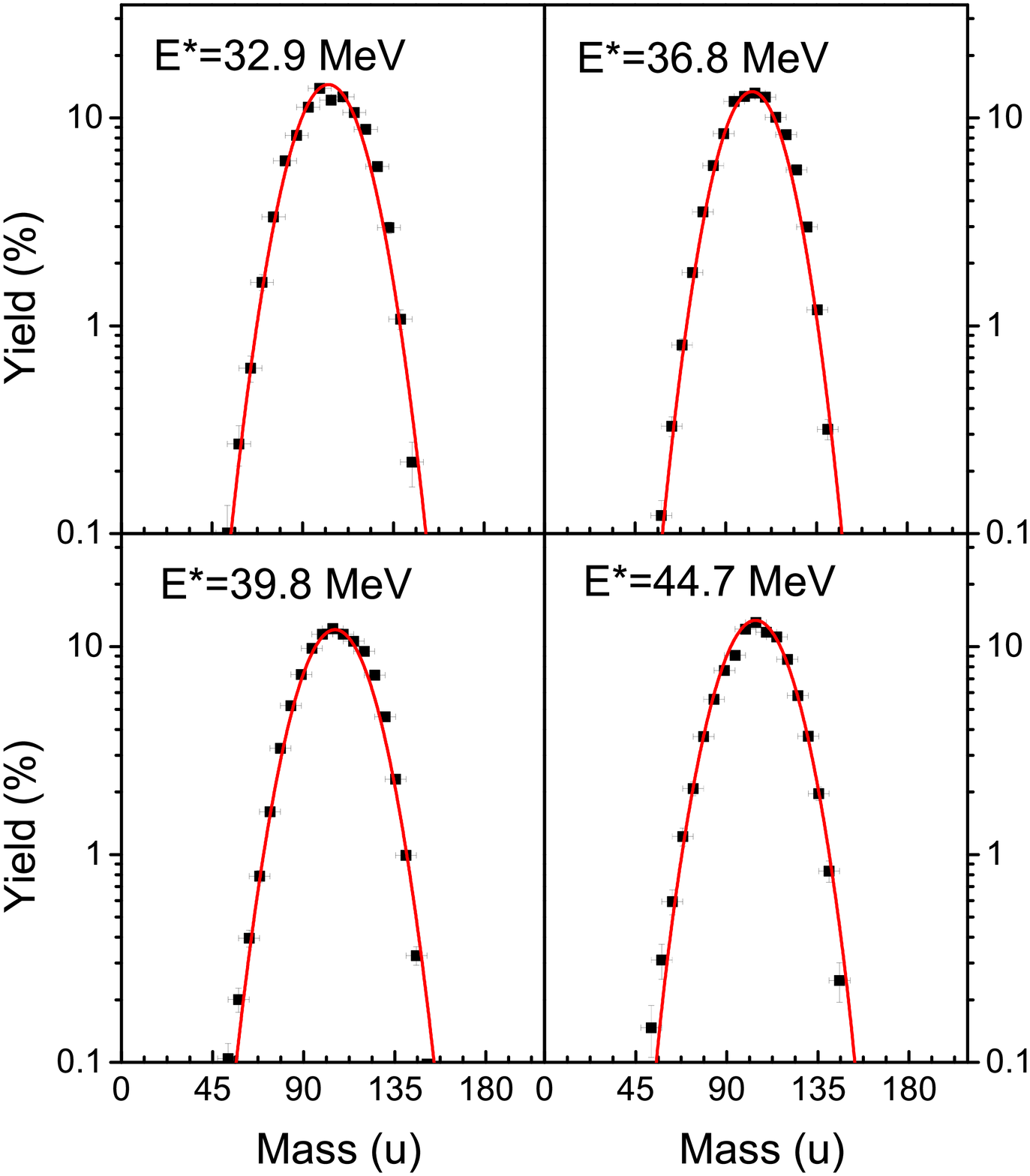}
\caption{\label{fig:fig4}~(Color online) Fission fragment mass distribution for $^{213}$At at various excitation energies. The red lines show the single Gaussian fit.}
\end{figure}

We have also studied the fission mass distribution  of the neighboring nucleus $^{213}$At produced in the reaction $^{4}$He + $^{209}$Bi for similar excitation energies, which is shown in figure 4. For this system also the mass distributions were found to be symmetric and could be fitted with a single Gaussian in the entire range of our measurement (E* $\approx$ 32 - 45 MeV). For this system, a weak asymmetric component was predicted at E* $\approx$ 25.69 MeV \cite{Moller2015}, which however, could not be reached due to lower cross sections. From the present study it can only be inferred that the shell effect in $^{213}$At is not prominent above E* $\approx$ 32 MeV; measurement at lower excitation energy $\approx$ 25 MeV is needed to verify if shell effect persists at that energy as predicted by the theory.

\begin{figure}
\includegraphics*[scale=0.35, angle=0]{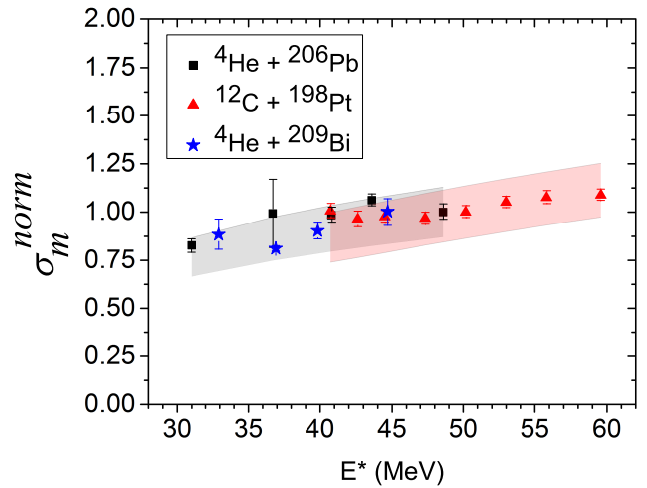}
\caption{\label{fig:fig5}~(Color online)The variation of the width of the mass distribution with excitation energy. The shaded region shows the uncertainty of the statistical model values.}
\end{figure}
 
In order to further characterise the origin  of the observed mass distributions, we now concentrate on the widths of the mass distributions. The statistical model fission mass width is represented by the following expression,

\begin{equation}
\sigma_{m}(u)= \sqrt{\alpha T + \beta<l^{2}>}
\end{equation}

where, T is the saddle point temperature and $<l^{2}>$ is the mean squared angular momentum of the fissioning system.  It is clear from the above equation that in case of pure statistical fission (without any shell effect), the width $\sigma_{m}$  increases only slowly and smoothly with T (or E*). On the other hand, for any change in the reaction mechanism (onset of quasi-fission and/or shell effect), there is abrupt change in the value of $\sigma_{m}$  \cite{AbhirupPRC15}. In figure 5, we have plotted the variation of the width of fragment mass distribution with excitation energy for $^{210}$Po populated using two different entrance channels: the present measurement ($^{4}$He + $^{206}$Pb)  and our earlier measurement ($^{12}$C + $^{198}$Pt) \cite{AbhirupRapid}. In the same graph, we have also plotted the fission fragment widths for $^{213}$At, populated through the reaction  $^{4}$He + $^{209}$Bi for comparison. To take care of the variation of  systematic errors in different measurements, the normalised values of $\sigma_{m}^{norm}$ (= $\sigma_{m}$ ($E^*$) / $\sigma_{m}$($E_{0}^*$))  have been plotted as a function of E*. The value of $E_{0}^*$ is suitably chosen at 50 MeV where there is no shell effect. It is seen that the normalised widths for all systems, irrespective of the entrance channel, follow identical trend. The widths for $^{4}$He + $^{206}$Pb reaction are found to be slightly higher than that of $^{12}$C + $^{198}$Pt reaction due to the higher average angular momentum $<l^{2}>$ (calculated using the code CCFUS \cite{CCFUS}) in the former reaction. However, it is seen that all points lie within the phenomenological limits of the statistical model predictions (shown as shaded bands) \cite{Galina1}, confirming the absence of shell effect in any of the cases above.

\maketitle{
\begin{center}
\textbf{Conclusions}
\end{center}}

To sum up, fission fragment mass distribution in reactions $^{4}$He + $^{206}$Pb populating $^{210}$Po nuclei has been measured over excitation energy range of $\approx$ 30 - 50 MeV. We observe slightly asymmetric mass distribution at 30.8 MeV of excitation energy indicating the preference for asymmetric saddle to scission pathway (persistence of shell effect). This is consistent with the recent theoretical description of the saddle to scission path in terms of both macroscopic-microscopic \cite{Moller2015} and self consistent \cite{McDonnell} approaches. Fragment mass distribution in reactions $^{4}$He + $^{206}$Pb populating $^{213}$At nuclei showed predominately symmetric mass distribution in the entire range of our measurement (excitation energy 32.9 - 44.7 MeV), which is also in agreement with the above theoretical predictions \cite{Moller2015}. The fragment mass widths for all systems studied, irrespective of the entrance channel, showed identical trend and their values were found to be within the limits of phenomenological statistical model predictions. Thus, the observed symmetric mass distributions at all excitation energies above $\approx$ 35 MeV  reaffirms the absence of shell effect at higher excitation. 

In conclusion, the present measurement confirmed the existence of shell effect vis-a-vis asymmetric fission pathway in $^{210}$Po at 31 MeV excitation, as predicted by the theory. Similar studies for other pre-actinides at low excitations are needed to have a proper understanding of the interplay of shell effect and dynamics in fission in this interesting Hg-Pb region.

\maketitle{
\begin{center}
\textbf{Acknowledgments}
\end{center}

The authors are thankful to VECC Cyclotron staff for providing high quality beams for the experiment. One of the authors (S. B.) acknowledges with thanks the financial support received as Raja Ramanna Fellow from the Department of Atomic Energy, Government of India.

\end{document}